\documentstyle[12pt]{article}

\def\beq{\begin{equation}}
\def\eeq{\end{equation}}
\def\bea{\begin{eqnarray}}
\def\eea{\end{eqnarray}}
\def\beas{\begin{eqnarray*}}
\def\eeas{\end{eqnarray*}}
\def\nn{\nonumber}
\begin{document}

\begin{titlepage}

\begin{flushright}{HU-TFT-91-44\\FIAN/TD/04-91\\ITEP-M-4/91}\end{flushright}
\begin{center}
\vspace{0.1in}{\Large\bf  On Equivalence of Topological\\
 and Quantum 2d Gravity}\\[.4in]
{\large  A.Marshakov, A.Mironov}\\
\bigskip {\it Research Institute for Theoretical Physics\\
 University of Helsinki\\ Siltavuorenpenger 20 C,
 SF-00170 Helsinki, Finland}\\{\large and}\\
 {\it Department of Theoretical Physics \\  P.N.Lebedev Physical
Institute \\ Leninsky prospect, 53, Moscow, 117 924}
\footnote{E-mail address: theordep@sci.fian.msk.su},\footnote{Permanent
address}\\ \smallskip
\bigskip {\large A.Morozov}\\
 \bigskip {\it Research Institute for Theoretical Physics\\
 University of Helsinki\\ Siltavuorenpenger 20 C, SF-00170
 Helsinki, Finland}\\{\large and}\\
{\it Institute of Theoretical and Experimental
Physics \\
 Bol.Cheremushkinskaya st., 25, Moscow, 117 259}
\footnote{Permanent
 address}
\end{center}

\newpage
\vskip .3in
\centerline{\bf ABSTRACT}
\begin{quotation}

We demonstrate the equivalence of Virasoro constraints imposed on continuum
limit of partition function of Hermitean 1-matrix model and the Ward identities
of Kontsevich's model. Since the first model describes ordinary  $d = 2$
quantum gravity, while the second one is supposed to coincide with Witten's
 topological
gravity, the result provides a strong implication that the two models are
indeed the same.

\end{quotation}
\end{titlepage}

\setcounter{page}2
\setcounter{footnote}0

\section{Introduction}

 From the early days of matrix models there is a belief that
partition functions of two {\it a priori} different models of 2d gravity should
coincide. The first is the (square root of) partition function
$\sqrt{Z_{qg}\{T_n}\}$  of ordinary Polyakov's 2d {\it quantum gravity},
 described by the
continuum limit of Hermitean 1-matrix model [1] (and is known to be a
$\tau $-function of KdV hierarchy [2]). The second partition function  $Z_{tg}$
of Witten's {\it topological gravity} [3]
 is {\it a priori} defined as a kind of generating
functional of intersection indices of divisors on module spaces of Riemann
surfaces with punctures.

Later it was discovered [4,5] that  $\sqrt{Z_{qg}\{T_n}\}$  satisfies a set of
Virasoro constraints:

\begin{equation}
{\cal L} _n\sqrt{Z_{qg}\{T}\} = 0\hbox{, }   n \geq  -1
\end{equation}

\begin{eqnarray*}
{\cal L} _n = \sum _{k\geq \delta_{n+1,0} }
(k+{1\over 2})
T_k {\partial \over \partial T_{k+n}} + \sum _{^{a+b=n-1}_{a,b\geq 0}}
{\partial ^2\over \partial T_a\partial T_b} +
\delta _{n+1,0}\cdot {T^2_0\over 16} + \delta _{n,0}\cdot {1\over 16}
\end{eqnarray*}

Eq. (1) can be deduced [6] as a continuum limit of Ward identities in discrete
matrix model, associated with the shift of integration variables [7]

\beq
X  \rightarrow   X + \epsilon X^{p+1}
\eeq
in the integral over $N \times N$ Hermitean matrix

\beq
Z^{(d)}_N\{t\} = \int   DX\ \exp  - trV\{X\}
\eeq

\begin{eqnarray*}
V\{X\} = \sum ^\infty _{k=0}t_kX^k
\end{eqnarray*}
However, it may be more reasonable to take these Virasoro constraints for a
straightforward definition of  $Z_{qg}\{T\}$, which does not refer to a
sophisticated change of variables  $\{t\} \rightarrow  \{T\}$ [6] and to
discrete model (3) at all.

As for  $Z_{tg}$ it was recently represented by M.Kontsevich [8] in terms of
another matrix model

\beq
Z_{tg} = \lim_{size {X}{\rightarrow }
_\infty}  Z^{(d)}_{tg}
\eeq
where

\beq
Z^{(d)}_{tg} = {1\over C[M]} \int   DX\ \exp  - tr\{MX^2 + X^3\}
\eeq
with

\beq
C[M] = \int   DX\ \exp  - tr\{MX^2\} = \det (M\otimes I +
I\otimes M)^{-{1\over2}}
\eeq
and  $X,M$  being (anti)Hermitean matrices.

It is a simple combinatorial result, that as soon as the size of  $X$  goes to
infinity  $Z^{(d)}_{tg}$ becomes dependent only on the variables
\footnote{We shall see the origin of  $\delta _{m,1}$ term and the
${1\over m+{1\over 2}}$  factor below. In these both aspects (7) is different
from Kontsevich's definition of times variables. Though his choice is more
appropriate for the study of generating functional for intersection indices,
(7) is better to stablish the correspondence with 1-matrix model.
In the other words, the generating functional of intersection indices appears
to be a $\tau-$function only after appropriate rescaling of "time"-variables.
}

\beq
T_m = {{3^{2m+1}}\over m+{1\over 2}} tr\ M^{-2m-1} - {4\over 3\sqrt{3}}
 \delta _{m,1}
\eeq

The explicit formulation of the original suggestion in these terms would be

\beq
Z_{tg}\{T_m\} = \sqrt{Z_{qg}\{T_n}\}
\eeq
which in particular implies the what is known as Witten's suggestion,
 $i.e$. that
$Z_{tg}\{T\}$  is like  $\sqrt{Z_{qg}\{T}\}$  a  $\tau $-function of
KdV-hierarchy. A necessary condition for (8) to hold is that  $Z_{tg}\{T\}$
defined by (4-6) satisfies the same Virasoro constraints (1).

This is the statement which we are going to prove in the paper. Moreover we
shall prove the relation involving  $Z^{(d)}_{tg}$ (see eq.(16) below), $i.e$.
valid for {\it finite} dimensional matrices $X$, which implies the entire set
of Virasoro constraints only as the size goes to infinity.

This statement is equivalent to (8) modulo:

(i) the assertions which have been made by Kontsevich in the derivation of
(4-6) from Penner's formalism of fat graphs, and

(ii) the so far subtle problem of uniqueness of solutions to Virasoro
constraints (1).
\bigskip

We know at least about two recent attempts to develop some similar $(i.e$.
exploiting Virasoro constraints) approach to the study of suggestion (8). One
is due to E.Witten [9]
\footnote{We were informed about this talk by prof. I.Singer, but
 unfortunately we
do not know what exactly has been done.
}. Another one is due to Yu.Makeenko and G.Semenoff [10] who investigated the
relation between tree level approximation to  $\sqrt{Z_{qg}}$  and  $Z_{tg}$.
Explicit relation between [10] and our reasoning below deserves further
investigation.

\section{Main results}

Since the Ward identities are a little obscure in Kontsevich's
presentation (4-6), we begin with a slight reformulation of his model.

After a shift of integration variable  $X  \rightarrow   X - {M\over 3}$  and
the redefinition  $M^2 = 3\Lambda $, we obtain

\beq
{\cal F}\{\Lambda \} \equiv  \int   DX\ \exp (- trX^3 + tr\Lambda X)
= C[\sqrt{\Lambda }] \exp ({2\over 3\sqrt{3}}tr\Lambda ^{3/2}) Z^{(d)}_{tg}
\eeq
with

\beq
C[\sqrt{\Lambda }] = \det (\sqrt{\Lambda }\otimes I +
I\otimes \sqrt{\Lambda })^{-{1\over2}}
\eeq

The functional  ${\cal F}\{\Lambda \}$  can be considered as a kind of
matrix-Fourier transform of exponential cubic potential  $trX^3$. It satisfies
obvious Ward identities, associated with a shifts of integration variables

\beq
X  \rightarrow   X + \epsilon _p
\eeq
namely

\beq
tr(\epsilon _p {\partial ^2\over \partial \Lambda ^2_{tr}} -
{1\over 3}\epsilon _p\Lambda ){\cal F}\{\Lambda \} = 0
\eeq
Here  $\epsilon _p$ stands for any $X$-independent {\it matrix}, and
$\Lambda _{tr}$ denotes transponent matrix. There is as many as $N^2$ different
choices of  $\epsilon _p$  ($N$ being the size of $X$ matrix). Though the
 concrete
choice of  $\epsilon _p$ should not be essential, we shall assume it to be a
function of $\Lambda $, $e.g$.  $\epsilon _p = \Lambda ^p$ (with integer or
half integer  $p$  , again it is not essential). In principle, there is nothing
to forbid consideration of  $X$-dependent  $\epsilon _p's$, for example
$\epsilon _p = X^{p+1}$ (as implied by [7]). This case was investigated to some
extent in [10], but it leads to much more complicated Ward identities than (12)
(because the integration measure is no longer invariant, and, what is more
important, the cubic form of potential is not preserved). Clearly, the complete
basis  $\epsilon _p = X^{p+1}$ is somehow related to the basis  $\epsilon _p =
\Lambda ^p$ and after all the Ward identities should be essentially the same
(indeed, they were interpreted in [10] as Virasoro constraints), but it is not
so simple to prove \footnote{In fact, in actual derivating in ref.10 some
other "master equations" are used (but not derived) instead of these Virasoro
constraints. Master equation (eq.(3.1) in ref.10) is certainly nothing but
our Ward identity (12) with $(\epsilon )_{ij} = \delta_{pi} \delta_{pj}$
  ($p = 1 \ldots N$) in the basis, where $\Lambda$ is diagonal (comp. with (20)
below).}.

Let us return to our Ward identities (12). Substitute in (12) the functional
${\cal F}$  in the form, inspired by (9)

\beq
{\cal F}\{\Lambda \} = C[\sqrt{\Lambda }] \exp (
{2\over 3\sqrt{3}}tr\Lambda ^{3/2}) Z\{T_m\}
\eeq
with

\beq
T_m = {1\over m+{1\over 2}} tr\Lambda ^{-m-{1\over2}} -
{4\over 3\sqrt{3}}\delta _{m,1}
\eeq
Our main statement (to be proved in sect.3 below) is that after the
substitution of (13) into (12) it turns into

\beq
\sum _{n\geq -1}tr(\epsilon _p\Lambda ^{-n-2}) {\cal L} _nZ = 0
\eeq
The identity

\beq
{1\over {\cal F}} tr(\epsilon _p {\partial ^2\over \partial \Lambda ^2_{tr}} -
{1\over 3}\epsilon _p\Lambda ){\cal F} = {1\over Z}
\sum _{n\geq -1} {\cal L}_nZ \hbox{   } tr(\epsilon _p\Lambda ^{-n-2})
\eeq
is valid {\it for any size of the matrix $\Lambda $}, but only in the limit of
infinitely large $\Lambda :$

(i) it is reasonable to substitute $Z^{(d)}_{tg}$ in (9) by $Z_{tg}$, which
depends only on  $tr\Lambda ^{-q}$ with half-integer $q's;$

(ii) the $T_m$'s in (14) are really independent variables;

(iii) all the quantities

\beq
tr(\epsilon _p\Lambda ^{-n-2})
\eeq
$(e.g$. $tr\Lambda ^{p-n-2})$ become algebraically independent, so that eq.
(15) implies that

\beq
{\cal L} _nZ_{tg}\{T\} = 0\hbox{, }    n \geq  -1
\eeq
\bigskip

This concludes our derivation of Virasoro constraints for  $Z_{tg}\{T\}$
defined as continuum limit of Kontsevich's partition function.

Note that the fact that operators  ${\cal L} _n$ in (15) contain only second
$T$-derivatives, is a direct consequence of that only double
$\Lambda $-derivatives arise in the $l.h.s$. and thus of the cubic nature of
potential  $trX^3$ in Kontsevich's model. It is very interesting (in
particular, from the point of view of multimatrix and Potts models) to study
the matrix-Fourier transform of potentials like  $trX^{K+1}$ for any degree
$K+1$. Amusingly enough, the corresponding Ward identities being transformed to
the form of (15) contain operators similar to the higher-spin operators of the
$W_K$-algebra instead of the Virasoro generators  ${\cal L}_n$. This should be
somehow related to the recent suggestion of E.Gava and K.Narain [11], namely
that the Ward identities in continuum limit of 2-matrix model acquire the form
of $W_K$-algebra constraints if one of potentials is of degree  K. For a more
close relation between [11] and the Ward identity (12) see [12].

\section{The proof of the main result}
\subsection{An important relation}

First, let us point out that ${\cal F} \{\Lambda \}$ defined by (13) which
we have to differentiate to prove (16) depends only upon the eigenvalues
$\{\lambda _k\}$ of the matrix $\Lambda $. Therefore, it is natural to consider
eq.(16) at the diagonal point $\Lambda _{ij}=0$, $i\neq j$. The only
``non-diagonal" piece of (16) which survives at this point is proportional to

\beq
\left.{\partial ^2\lambda _k\over \partial \Lambda _{ij}\partial \Lambda _{ji}}
\right| _{\Lambda _{mn}=0\hbox{, } m\neq n} =
{\delta _{ki}-\delta _{kj}\over \lambda _i-\lambda _j}\hbox{
for }\ i\neq j.
\eeq

Eq.(19) is nothing but a familiar formula for the second order correction to
 Hamiltonian
eigenvalues in ordinary quantum-mechanical perturbation theory. It can be
easily derived from the variation of determinant formula:

\begin{eqnarray*}
\delta log(\det \ \Lambda ) = tr {1\over \Lambda } \delta \Lambda  - {1\over 2}
tr( {1\over \Lambda } \delta \Lambda  {1\over \Lambda } \delta \Lambda )
 + \ldots \hbox{ .}
\end{eqnarray*}
For diagonal $\Lambda _{ij}= \lambda _{i}\delta _{ij}$, but, generically,
non-diagonal $\delta \Lambda _{ij}$, this equation gives

\begin{eqnarray*}
\sum  _k {\delta \lambda _k\over \lambda _k} = - {1\over 2} \sum _{i\neq j}
{\delta \Lambda _{ij}\delta \Lambda _{ji}\over \lambda _i\lambda _j} =
{1\over 2} \sum _{i\neq j} \left( {1\over \lambda _i} -
{1\over \lambda _j}\right)
{\delta \Lambda _{ij}\delta \Lambda _{ji}\over \lambda _i-\lambda _j} +
\ldots \hbox{ ,}
\end{eqnarray*}
which proves (19).

If applied to a function ${\cal F} \{\Lambda \}$ depending only on eigenvalues
of $\Lambda $, eq.(19) implies that$:$

\bea
\left( {\partial ^2\over \partial \Lambda ^2}\right) _{ii}{\cal F}
\left.\{\Lambda \}\right| _{\Lambda _{mn}=0\hbox{, } m\neq n} = \sum  _j
{\partial ^2{\cal F}
\{\Lambda \}\over \partial \Lambda _{ij}\partial \Lambda _{ji}}
\left. \right| _{\Lambda _{mn}=0\hbox{, } m\neq n} = \nn \\
= \sum  _j
{\partial {\cal F}\ /\partial \lambda _i-\partial {\cal F}\ %
/\partial \lambda _j\over \lambda _i - \lambda _j}
\eea
(note that the item $j=i$ is not omitted from the sum).

The general formula for matrix derivative of such functional at diagonal point
is:

\bea
\left( {\partial ^{n+1}\over \partial \Lambda ^{n+1}}\right) _{ii}{\cal F}
\left.\{\Lambda \}\right| _{\Lambda _{mn}=0\hbox{, } m\neq n} = \nn \\
= \sum _{j_1 \ldots j_n} \hbox{   } \sum _{^{over \hbox{   }
 permutations}_{[i,j_1 \ldots j_n]}}
{\partial {\cal F}\ /\partial \lambda _i\over (\lambda _1 -
\lambda _{j_1} ) \ldots (\lambda _i -
\lambda _{j_k} )}\hbox{ . }
\eea

\subsection{Simplified example}

In order to illustrate the idea, we shall begin now with a simpler calculation
which is (i) much more transparent that actual derivation of (16), and (ii) can
be also of interest from the point of view of relation between discrete and
continuum matrix models. Namely, let us derive the formula for the action of
$tr\{\epsilon _p\partial ^2/\partial \Lambda ^2\}$ on a function
$Z^{(d)}\{t_k\}$, $t_k= {1\over k} tr {1\over \Lambda ^k}$, $t_0= -
tr\{log\Lambda \}$ (in variance with (14) the integer powers of $\Lambda $ are
involved). Then (using (19))

\bea
tr\left( \epsilon _p
{\partial ^2\over \partial \Lambda ^2_{tr}}\right) Z(t)& = & \sum _{k,l\geq
0}{%
\partial ^2Z\over \partial t_k\partial t_l}tr\left( \epsilon _p\Lambda ^{-k-l-2
}\right)  +\sum _{n\geq 0}{\partial Z\over \partial t_n}tr\left( \epsilon _p
{\partial ^2t_n\over \partial \Lambda ^2_{tr}}\right)= \nn \\
& = & \sum _{n\geq -1} {\cal L}_nZ \hbox{   }
tr\left( \epsilon _p\Lambda ^{-n-2}\right)
\eea
with

\beq
{\cal L}_n= \sum _{k\geq 1}kt_k {\partial \over \partial t_{k+n}} +
\sum _{^{a+b=n}_{a,b\geq 0}}{\partial ^2\over \partial t_a\partial t_b}%
\hbox{ .}
\eeq
Remarkably enough, the operators (23) are exactly the Virasoro operators
annihilating partition function (3) of discrete matrix model. In these models
also the $t_0$-dependence of $Z^{(d)}$ is fixed:

\begin{eqnarray*}
{\partial \over \partial t_0} Z^{(d)} = -NZ^{(d)}.
\end{eqnarray*}
This corresponds to extracting the factor  $e^{-Nt_0} = (det\Lambda )^{-N}$
 from $Z^{(d)}$. This factor should be compared with

\begin{eqnarray*}
C(\sqrt{\Lambda }) = [\det (\sqrt{\Lambda } \otimes  I + I \otimes
\sqrt{\Lambda })]^{-1/2}
\end{eqnarray*}
in the continuum case. Note that while  $C(M) = [\det (M \otimes  I + I
\otimes  M)]^{-1/2}$ is a normalization factor for Gaussian measure
$dX\exp \{- tr(MX^2)\}$, in the discrete case we have

\beq
(det\Lambda )^{-N} = det ^{-1/2} M^N = det ^{-1/2}(\sqrt{M} \otimes
\sqrt{M}).
\eeq
This represents a normalization factor for slightly different Gaussian measure
, that is,
$dX\ \exp \{- tr(\sqrt{M}X\sqrt{M}X)\}.$

The derivation in this subsection is not of immediate use, since (i) we did not
find anything like Ward identity

\begin{equation}
tr\left( \epsilon _p {\partial ^2\over \partial \Lambda ^2_{tr}}\right) Z(t) =
0,
\end{equation}
and (ii) in the case of finite size $N$ of the matrices there is no way to
distinguish particular terms in the sum over $n$ in (22) (the quantities
$tr\{\epsilon _p\Lambda ^{-n-2}\}$ are algebraically dependent). Still it is an
amusing reformulation of discrete Virasoro constraints, especially because it
is somewhat generalizable:

\beq
tr \epsilon _p \left( {\partial ^n\over \partial \Lambda^n_{tr}
}\right)  Z^{(d)}\{t\} = \sum _{n\geq 1-K} \tilde W^{(K)}_n Z^{(d)}\{t\}
\hbox{   } tr\left( \epsilon _p \Lambda ^{-K-n} \right )
\eeq
is similarly represented as a sum of harmonics of the (discrete-model)
$\tilde W^{(K)}$-operators,

\bea
\tilde W^{(K)}_n & = & \sum_{k_1, \ldots ,k_{K-1}}
    k_1t_{k_1}...k_{K-1}t_{k_{K-1}}
{\partial \over \partial t_{k_1+ \ldots + k_{K-1}+n}} + \ldots + \nn \\
& + & \sum _{a_1+ \ldots +a_K=N}
{\partial ^K\over \partial t_{a_1}...\partial t_{a_K}}\hbox{ .}
\eea
Tilde over $\tilde W$ stands to emphisize that the relative coefficients in the
sum (27) do not coinside with the conventional definition of
$W^{(K)}$-operators of $W$-algebra. While (16) ---
 the continuum-model analogue of (22) ---
will be explicitely derived in $s.3.3$. below, the analogue of (27), which
hopefully exists as well, deserves more efforts to be worked out.

\subsection{Derivation of eq.(16)}

Now we shall turn directly to the proof of (16). Since $\epsilon _p$ is assumed
to be a function of $\Lambda $, it can be, in fact, treated as a function of
eigenvalues $\lambda _i$. After that, (16) can be rewritten in the following
way:

\beq
{e^{-{2\over 3\sqrt 3}tr\Lambda ^{3/2}}\over C(\sqrt \Lambda )Z\{T\}}\left[ tr\
\epsilon _p\{{\partial ^2\over \partial \Lambda ^2} - {1\over 3}
\Lambda \}\right]  C(\sqrt \Lambda ) e^{
{2\over 3\sqrt 3}tr\Lambda ^{3/2}}Z\{T\} =
\eeq

\beq
= {1\over Z}\sum _{a,b\geq 0}{\partial ^2Z\over \partial T_a\partial T_b}
\sum  _i \epsilon _p(\lambda _i)
{\partial T_a\over \partial \lambda _i}\cdot {\partial T_b\over %
\partial \lambda _i} +
\eeq

\bea
+ {1\over Z} \sum _{n\geq 0}{\partial Z\over \partial T_n} \left[ \sum _{i,j}
\epsilon _p(\lambda _i)
{\partial ^2T_n\over \partial \Lambda _{ij}\partial \Lambda _{ji}} + 2 \sum  _i
\epsilon _p(\lambda _i) {\partial T_n\over \partial \lambda _i}
{\partial logC\over \partial \lambda _i} +\right. \nn \\
+ 2 \sum  _i \epsilon _p(\lambda _i) {\partial T_n\over \partial \lambda _i}
\left( {2\over 3\sqrt 3}\right)  {\partial \over \partial \lambda _i}
tr\Lambda ^{3/2}\left. \right]  +
\eea

\beq
+ \left[
 \sum  _i
\epsilon _p(\lambda _i)
\left( {\partial \over \partial \lambda _i}\left\lbrace
{2\over 3\sqrt 3}\right\rbrace  tr\Lambda ^{3/2}\right) ^2 \right.
- {1\over 3} \sum  _i \lambda _i \epsilon _p(\lambda _i) +
\eeq

\beq
+ \sum _{i,j}\epsilon _p(\lambda _i)
\left( {\partial ^2\over \partial \Lambda _{ij}\partial \Lambda _{ji}}%
\left\lbrace {2\over 3\sqrt 3}\right\rbrace  tr\Lambda ^{3/2}\right)  +
\eeq

\beq
+ 2 \sum  _i \epsilon _p(\lambda _i) \left\lbrace
{2\over 3\sqrt 3}\right\rbrace
{\partial tr\Lambda ^{3/2}\over \partial \lambda _i}
{\partial logC\over \partial \lambda _i} +
\eeq

\beq
+ {1\over C} \sum _{i,j}\epsilon _p(\lambda _i)
{\partial ^2C\over \partial \Lambda _{ij}\partial \Lambda _{ji}}\left. \right]
\eeq
with  $tr\Lambda ^{3/2} = \sum  _k \lambda ^{3/2}_k$ and

\beq
C = \prod _{i,j}(\sqrt{\lambda }_i + \sqrt{\lambda }_j)^{-1/2}.
\eeq

First, since

\beq
{\partial T_n\over \partial \lambda _i} = - \lambda ^{-n-3/2}_i
\eeq
it is easy to notice that the term with the second derivatives (29) can be
immediately written in the desired form:

\bea
(29) =\sum _{n\geq -1}\sum  _i \epsilon _p(\lambda _i) \lambda ^{-n-2}_{i}
\sum _{a+b=n-1}{\partial ^2Z\over \partial T_a\partial T_b}= \nn \\
=\sum _{n\geq -1
}tr\{\epsilon _p(\lambda _i)\Lambda ^{-n-2}\}\sum _{a+b=n-1}{\partial ^2Z%
\over \partial T_a\partial T_b}\hbox{.}
\eea
For the first-derivative terms with the help of (35),

\beq
{\partial logC\over \partial \lambda _i} = - {1\over 2\sqrt \lambda _i} \sum
_j {1\over \sqrt \lambda _i+\sqrt \lambda _j}
\eeq

\beq
{\partial \over \partial \lambda _i} tr\Lambda ^{3/2} = {3\over 2}
\sqrt{\lambda }_i
\eeq
and

\beq
{\partial ^2T_n\over \partial \Lambda _{ij}\partial \Lambda _{ji}} = \sum  _k
{\partial ^2\lambda _k\over \partial \Lambda _{ij}\partial \Lambda _{ji}}
{\partial T_n\over \partial \lambda _k} +
{\partial ^2T_n\over \partial \lambda^2_i}
 \delta _{ij}\hbox{ .}
\eeq
Then we have

\bea
{1\over Z} \sum _{n\geq 0} {\partial Z\over \partial T_n}
\left[ \sum _{i,j}\left\lbrace \epsilon _p(\lambda _i) {\lambda_ i^{n-3/2}
\lambda_j^{n-3/2}\over \sqrt \lambda _i+\sqrt \lambda_j}
+\sum _{a+b=2(n+1)}(\sqrt{\lambda }_i)^a(\sqrt{\lambda }_j)^b
+\right. \right. \nn \\
+ \lambda ^{-n-2}_i {1\over \sqrt \lambda _i+\sqrt \lambda _j}
\left. \right\rbrace  - {2\over \sqrt 3} \sum  _i \epsilon _p(\lambda _i)
\lambda ^{-n-1}_i \left. \right]  = \nn \\
= {1\over Z} \sum _{n\geq 0}{\partial Z\over \partial T_n} \left[ \sum _{i,j}
\sum ^{n-1}_{a=0}\epsilon _p(\lambda _i) \lambda ^{-n-2+a}_i
\lambda ^{-a-1/2}_j - \sum  _i {2\over 2\sqrt 3}
\epsilon _p(\lambda _i)\lambda _i^{-n-1}\right]  = \nn \\
= {1\over Z} \sum _{n\geq -1} \sum  _i \epsilon _p(\lambda _i)
\lambda ^{-n-2}_{i} \sum _{k\geq \delta _{n+1,0}}\left( \sum  _j
\lambda ^{-k-1/2}_j - {2\over \sqrt 3} \delta _{k,1}\right)
{\partial Z\over \partial T_{n+k}}\left. \right]  =  \nn \\
= {1\over Z} \sum _{n\geq -1}tr(\epsilon _p\Lambda ^{-n-2}) \sum _{k\geq
 \delta _{n+1,0}}(%
k + {1\over 2}) T_k {\partial Z\over \partial T_{n+k}},
\eea
where

\begin{eqnarray*}
T_k = {1\over k+1/2} tr\Lambda ^{-k-1/2} - {4\over 3\sqrt 3} \delta _{k,1}.
\end{eqnarray*}

The remaining part contains terms proportional to $Z$ itself which need a bit
more care. First, it is easy to notice that two items in (31) just cancell each
other, so the (31) gives no contribution to the final result. For (32) we have

\beq
 {2\over 3\sqrt 3} \left[ \sum  _i \epsilon _p(\lambda _i)
{\partial ^2tr\Lambda ^{3/2}\over \partial \lambda^2_i
} + \sum _{i,j}\epsilon _p(\lambda _i)
{\partial tr\Lambda ^{3/2}\over \partial \lambda _k}
{\partial ^2\lambda _k\over \partial \Lambda _{ij}\partial \Lambda _{ji}}%
\right] \hbox{ . }
\eeq

Using (39) and (19) it can be transformed into

\bea
 {1\over \sqrt 3} \left[ {1\over 2} \sum  _i \epsilon _p(\lambda _i)
\lambda ^{-1/2}_i + \sum _{i\neq j}\epsilon _p(\lambda _i)
{\sqrt{\lambda }_i-\sqrt{\lambda }_j\over \lambda _i-\lambda _j}\right]
  = \nn \\
=  {1\over \sqrt 3} \sum _{i,j}\epsilon _p(\lambda _i)
{1\over \sqrt \lambda _i+\sqrt \lambda _j}
\eea
and this cancells (33) (using (38), (39)). Thus, the only contribution is (34),
which gives

\bea
{1\over C} {\partial ^2C\over \partial \Lambda _{ij}\partial \Lambda _{ji}} & =
& \sum  _k
{\partial ^2\lambda _k\over \partial \Lambda _{ij}\partial \Lambda _{ji}}
{\partial logC\over \partial \lambda _k} + {1\over C}
{\partial ^2C\over \partial \lambda^2_i
} \delta _{ij} = \nn \\
& = & \sum  _k {1\over 2(\lambda _i-\lambda _j)}
\left[ {1\over \sqrt \lambda _j(\sqrt \lambda _j+\sqrt \lambda _k)} -
{1\over \sqrt \lambda _i(\sqrt \lambda _i+\sqrt \lambda _k)}\right]  (1 -
\delta _{ij}) + \nn \\
& + & \delta _{ij}
\left[ {\partial ^2\hbox{logC}\over \partial \lambda^2_i
} + \left( {\partial logC\over \partial \lambda _i}\right) ^2\right] .
\eea

Now, using $(19')$ and (38) we obtain for (34):

\begin{eqnarray}
 & & \left\lbrace {1\over 16} \sum  _i \epsilon _p(\lambda _i)
\lambda ^{-2}_i +\right. \nn\\
& + &  {1\over 4} \sum _{i,j}\epsilon _p(\lambda _i) \lambda ^{-1}_i
\left( {1\over \sqrt \lambda _i(\sqrt \lambda _i+\sqrt \lambda _j)} +
{1\over (\sqrt \lambda _i+\sqrt \lambda _j)^2}\right)  + \nn \\
& + & {1\over 4} \sum _{i,j,k}\epsilon _p(\lambda _i) \lambda ^{-1}_i
{1\ \ \ \ \ \ \ \ 1\over (\sqrt \lambda _i+\sqrt \lambda _j)(\sqrt \lambda _i+%
\sqrt \lambda _k)} + \nn \\
& + & {1\over 2}\sum _{i\neq j,k}{\epsilon _p(\lambda _i)\over
 \lambda _i-\lambda _j
} \left[ {1\over \sqrt \lambda _j(\sqrt \lambda _j+\sqrt \lambda _k)} -
{1\over \sqrt \lambda _i(\sqrt \lambda _i+\sqrt \lambda _k)}\right] \left. %
\right\rbrace  =  \nn \\
& = & {5\over 16} \sum  _i \epsilon _p(\lambda _i) \lambda ^{-2}_i +
\sum _{i\neq j}\left( {1\over 4} \epsilon _p(\lambda _i)
\lambda ^{-1}_i\lambda ^{-1}_j + {1\over 2} \epsilon _p(\lambda _i)
\lambda ^{-3/2}_i\lambda ^{-3/2}_j\right)  + \nn \\
& + & \sum _{i\neq j\neq k}\epsilon _p(\lambda _i)\left( {1\over
 4\lambda _i(\sqrt %
\lambda _i+\sqrt \lambda _j)(\sqrt \lambda _i+\sqrt \lambda _k)} +\right. \nn
\\
& + & {1\over 2(\lambda _i-\lambda _j)}\left[\left.{1\over \sqrt
 \lambda _j(\sqrt \lambda
_j+\sqrt \lambda _k)} -
{1\over
 \sqrt \lambda _j(\sqrt \lambda _i+\sqrt \lambda _k)}\right] \right)  = \nn \\
& = & {5\over 16} \sum  _i \epsilon _p(\lambda _i) \lambda ^{-2}_i +
\sum _{i\neq j}\left( {1\over 4} \epsilon _p(\lambda _i)
\lambda ^{-1}_i\lambda ^{-1}_j + {1\over 2} \epsilon _p(\lambda _i)
\lambda ^{-3/2}_i\lambda ^{-3/2}_j\right)  + \nn \\
& + & {1\over 4}\sum _{i\neq j\neq k}\epsilon (\lambda )
\lambda ^{-1}_i\lambda ^{-1/2}_j\lambda ^{-1/2}_k.
\eea
In the last equality we used the fact that

\bea
\sum _{j\neq k}
{1\over (\lambda _i-\lambda _j)}\left[ {1\over \sqrt \lambda _j(\sqrt \lambda
_%
j+\sqrt \lambda _k)} -
{1\over \sqrt \lambda _j(\sqrt \lambda _i+\sqrt \lambda _k)}\right]  = \\
= {1\over 4} \sum _{j\neq k}
\left\lbrace {1\over (\lambda _i-\lambda _j)}\left[ {1\over \sqrt \lambda _j(%
\sqrt \lambda _j+\sqrt \lambda _k)} -
{1\over \sqrt \lambda _j(\sqrt \lambda _i+\sqrt \lambda _k)}\right]
\right. + \nn\\
+ \left.(j\leftarrow \rightarrow k)\right\rbrace \hbox{.} \nn
\eea

Finally, (45) can be rewritten as

\bea
{1\over 16} \sum  _i \epsilon _p(\lambda _i) \lambda ^{-2}_i + {1\over 4} \sum
_i \epsilon _p(\lambda _i) \lambda ^{-1}_i \sum _{j,k}
\lambda ^{-1/2}_j\lambda ^{-1/2}_k = \nn \\
= \sum  _n \left( \sum  _i \epsilon _p(\lambda _i)
\lambda ^{-n-2}_i\right) \left\lbrace {1\over 16} \delta _{n,-1}\left[ 2 \sum
_j \lambda ^{-1/2}_j\right] ^2 + {1\over 16} \delta _{n,0}\right\rbrace .
\eea

Now taking together (37), (41) and (47) we obtain our main result:

\bea
{e^{-{2\over 3\sqrt 3}tr\Lambda ^{3/2}}\over C(\sqrt \Lambda )Z\{T\}}\left[ tr\
\epsilon _p\{{\partial ^2\over \partial \Lambda ^2} - {1\over 3}
\Lambda \}\right]  C(\sqrt \Lambda ) e^{
{2\over 3\sqrt 3}tr\Lambda ^{3/2}}Z\{T\} = \nn \\
= {1\over Z} \sum _{n\geq -1}tr(\epsilon _p\Lambda ^{-n-2})
\left\lbrace \sum _{k\geq \delta _{n+1,0}}(k + 1/2) T_k
{\partial \over \partial T_k}
+\sum _{^{a+b=n-1}_{a\geq 0,b\geq 0}}{\partial ^2\over \partial T_a%
\partial T_b} +\right.\nn \\
+ {1\over 16} \delta _{n,0} + {1\over 16} \delta _{n+1,0} T^2_0
\left. \right\rbrace  Z(T) = 0.
\eea
This completes the derivation.

\bigskip
\bigskip
\begin{center}{{\Large\bf Aknowledgements}}\end{center}
\medskip

We are deeply indepted to prof.I.Singer who informed us about the talk [9] and
encouraged us to comlete our study of Ward identities in Kontsevich's model. We
benefited a lot from illuminating disscussions on the subject with E.Corrigan,
A.Gerasimov, S.Kharchev, M.Kontsevich, Yu.Makeenko, A.Niemi, I.Zakharevich.

We are grateful to the hospitality of Math. Sci. Res. Inst., Berkeley, CA and
the Research Institute for Theoretical Physics (TFT)
at Helsinki University were some pieces of this work have been done. A.Mor. is
also very much indepted for the hospitality of Dept. of Math. Sci. of Durham
University. The work of A.Mar. and A.Mir. was partially supported by the
Re\-se\-a\-rch In\-s\-ti\-tu\-te for The\-o\-re\-ti\-cal Physics (TFT) at
Hel\-sin\-ki Uni\-ver\-si\-ty while that of A.Mor. by the UK Science
 and En\-ge\-ne\-er\-ing
Co\-un\-cil through the Vi\-si\-ting Fel\-low\-ship prog\-ram and by
NO\-R\-DI\-TA.

\bigskip
\bigskip
\begin{center}{{\Large\bf References}}\end{center}
\medskip
1. V.Kazakov  Mod.Phys.Lett. A4 (1989) 2125

  E.Brezin, V.Kazakov   Phys.Lett. B236 (1990) 144

    M.Douglas, S.Shenker  Nucl.Phys. B335 (1990) 635

    D.Gross, A.Migdal  Phys.Rev.Lett. 64 (1990) 127\\
2. M.Douglas  Phys.Lett. B238 (1990) 176\\
3. E.Witten  Nucl.Phys. B340 (1990) 281

   R.Dijkgraaf, E.Witten  Nucl.Phys. B342 (1990) 486\\
4. M.Fukuma, H.Kawai, R.Nakayama  Int.J.Mod.Phys. A6 (1991) 1385\\
5. R.Dijkgraaf, E.Verlinde, H.Verlinde  Nucl.Phys. B356 (1991) 574\\
6. Yu.Makeenko et.al. Nucl.Phys. B356(1991) 574\\
7. J.Ambjorn, J.Jurkiewicz, Yu.Makeenko  Phys.Lett. B (1990)

    A.Gerasimov et.al. Nucl.Phys. B357 (1991) 565

    A.Mironov, A.Morozov Phys.Lett. B252 (1990) 47\\
8. M.Kontsevich  Funk.Anal. i Priloz. 25 (1991) 50\\
9. E.Witten  in talk at NYC conference, June 1991\\
10. Yu.Makeenko, G.Semenoff  ITEP/UBC preprint, July 1991\\
11. E.Gava, K.Narain  Phys.Lett. B263 (1991) 213\\
12. A.Marshakov, A.Mironov, A.Morozov preprint ITEP-M-5/91 --- FIAN/TD/05-91

\end{document}